\documentclass[
reprint,
superscriptaddress,
preprintnumbers,
amsmath,amssymb,
aps,
prl, 
nofootinbib]{revtex4-2}

\usepackage{graphicx}% Include figure files
\usepackage{braket}
\usepackage{dcolumn}% Align table columns on decimal point
\usepackage{bm}% bold math
\usepackage{xcolor}
\usepackage{slashed}
\usepackage{feynmf} % Load the feynmf package
\usepackage{cancel}
\begin{document}

\preprint{ANL-205269}

\title{Hadron Structure from the Hierarchy of Quantum Correlations in Deep-Inelastic Scattering}
%\title{Quantum Information as a Probe of Proton Substructure}
 % Force line breaks with \\

\author{Henry Bloss}
 \email{hbloss@hawk.illinoistech.edu}
\affiliation{%
Department of Physics, Illinois Institute of Technology, Chicago, IL 60616, USA}%

%3101 South Dearborn St., Chicago, IL 60616, United States}%
\affiliation{%
 High Energy Physics Division, Argonne National Laboratory, Lemont, IL 60439, USA}%

\author{T.J. Hobbs}
 \email{tim@anl.gov}
\affiliation{%
 High Energy Physics Division, Argonne National Laboratory, Lemont, IL 60439, USA}%
 \affiliation{%
Department of Physics, Illinois Institute of Technology, Chicago, IL 60616, USA}%

\author{Navin McGinnis}
 \email{nmcginnis@arizona.edu}
\affiliation{%
 Department of Physics, University of Arizona, Tucson, Arizona 85721, USA}%


%\affiliation{
%\affiliation{
%\affiliation{%


\date{\today}% It is always \today, today,

\begin{abstract}
We show that the hierarchy of quantum correlations produced in deep-inelastic scattering (DIS) can serve as a novel probe of the proton's nonperturbative structure. Specifically, we show that quantum entanglement, discord, steering, and magic provide nontrivial and complementary sensitivities to the nucleon's parton distribution functions (PDFs), particularly those encoding the transverse-spin polarization of the interacting quark. This connection leads to a unique probe of the proton's parton-level tensor charges with implications for beyond Standard Model (BSM) physics searches. We propose how quantum information measures can be utilized for precision studies of hadron structure at DIS experiments like the upcoming Electron-Ion Collider (EIC).
\end{abstract}

\maketitle

%\tableofcontents
%%%%%%%%%%%%%%%%%%%%%%%%%%%%%%%%%%%%%%%%%%%%%%%%%%%%%%%%%%%%%%%%%%%%%%%%%%%
\noindent\textbf{\textit{Introduction}} --- Understanding the partonic structure of the proton
has been a central problem in particle physics since the discovery of scaling~\cite{Bjorken:1968dy} in the earliest
deep-inelastic scattering (DIS) experiments. In addition to being a central problem in
modern Quantum Chromodynamics (QCD), detailed knowledge of the proton's internal quark-gluon
dynamics is essential for precision tests of the Standard Model (SM) and searches for possible 
beyond SM (BSM) physics in high-energy processes like Higgs production at the Large Hadron
Collider (LHC).
Contemporary information on proton substructure is encoded in parton distribution functions
(PDFs)~\cite{Hou:2019efy,Moffat:2021dji,Bailey:2020ooq,NNPDF:2021njg}, which probabilistically characterize the number densities of quarks and gluons inside
a hadron.
At leading twist, the collinear structure of the nucleon is captured by three
distributions: the unpolarized ($f_q$), helicity ($\Delta f_q$), and transversity ($h_{1,q}$) PDFs. Of
these, the transversity PDF~\cite{Kang:2015msa,Gamberg:2022kdb,Cocuzza:2023vqs} has been least constrained in QCD global analyses,
owing substantially to its chiral-odd nature: it cannot appear alone in an inclusive cross section, and must
instead be accessed in combination with another chiral-odd function, such as Collins or dihadron fragmentation functions.
Moreover, the lowest $C$-odd moments of $h_{1,q}$ define corresponding nucleon tensor charges, $\delta q$, which are independently calculable within
lattice-gauge theory and relevant for BSM searches, especially those pertaining to a possible neutron
electric dipole moment (EDM)~\cite{Chupp:2017rkp} or higher-dimensional operators entering $\beta$-decay. %~\cite{BSM_refs}. 
%

% A precise understanding of the transversity PDFs and of the tensor charges is
% therefore a crucial ingredient for precision physics involving the proton.
 
Current approaches to extracting transversity PDFs rely on measurements of azimuthal spin
asymmetries: the Collins effect in semi-inclusive deep-inelastic scattering (SIDIS), and
dihadron production in SIDIS and in proton-proton collisions, analyzed within global QCD fits
that simultaneously determine the relevant fragmentation
functions~\cite{Kang:2015msa,Gamberg:2022kdb,Cocuzza:2023vqs}. Because the available data constrain
only a finite range of intermediate parton fractions, the tensor-charge moments additionally require
extrapolation to very low and high $x$ ($\to 0, 1$); the combined parametrization and
extrapolation dependence and associated data sparsity drive a significant uncertainty and
underlies possible tension between phenomenological extractions and lattice QCD.
In this context, independent handles on transversity are highly desirable.
 
Recently, there has been a growing effort to study quantum information (QI) at present and
future colliders --- especially, its potential as a new tool for characterizing
fundamental particle interactions. Treating the spin of each particle as a qubit, the final
state of a scattering process is a multi-qubit quantum state, whose entanglement, quantum
discord, steering, and non-stabilizerness (``magic'') can be studied directly~\cite{Afik:2020onf,Fabbrichesi:2021npl,Severi:2021cnj,Aoude:2022imd,Fabbrichesi:2022ovb,Afik:2022dgh,Ashby-Pickering:2022umy,Severi:2022qjy,Altakach:2022ywa,Dong:2023xiw,Morales:2023gow,Aoude:2023hxv,Ma:2023yvd,Sakurai:2023nsc,Bernal:2023jba,Han:2023fci,Altomonte:2023mug,Ehataht:2023zzt,Maltoni:2024tul,Aguilar-Saavedra:2024hwd,Blasone:2024dud,Barr:2024djo,Subba:2024mnl,Bernal:2024xhm,CMS:2024pts,Wu:2024asu,Demina:2024dst,Gabrielli:2024kbz,Ruzi:2024cbt,CMS:2024zkc,Du:2024sly,Ravina:2024ard,Cheng:2024rxi,Sullivan:2024wzl,Wu:2024ovc,Ruzi:2024iqu,Altomonte:2024upf,Han:2024ugl,Fabbrichesi:2025ywl,Cheng:2025cuv,Lysak:2025uhk,Han:2025ewp,Fabbrichesi:2025aqp,Aoude:2025ovu,Fabbrichesi:2025psr,Goncalves:2025mvl,Aguilar-Saavedra:2025byk,Aoude:2025jzc,Afik:2025grr,Qi:2025onf,Goncalves:2025xer,Cheng:2025zcf,Bechtle:2025ugc,Abel:2025skj,Wu:2025dds,Aguilar-Saavedra:2025cej,Pei:2025ito,Gu:2025ijz,Yazgan:2025pah,Cheng:2025zaw,Cao:2025xnp,CMS:2025brx,Jolly:2026gpe,Pardos:2026uhw,Guo:2026yhz,Gabrielli:2026tnl,Zhang:2026nwm,Yang:2026uwu,Aguilar-Saavedra:2026wuq,Oussarhan:2026yli,Fang:2026ddi,Goncalves:2026njf,Aoude:2026eeg,Arai:2026jtc,ATLAS:2026nrx,Subba:2026nzs,Wang:2026nls,Zhang:2026wvn,Zhou:2026poo,Goncalves:2026nnx,Liu:2026gxj,Batell:2026bcd,Cheng:2026ktp,Lamba:2026sni,Lamba:2026jfm,Banacki:2026msu}. Notably, Ref.~\cite{Cheng:2025zaw} showed that transversely polarized
electron-proton scattering can generate nontrivial entangled and non-stabilizer final states at
facilities such as the Electron-Ion Collider (EIC)~\cite{AbdulKhalek:2021gbh}; in addition, QI principles have been shown~\cite{Benito-Calvino:2022kqa,Bloss:2025ywh,Galvez-Viruet:2025rmy,Hentschinski:2026otq} to provide insights into the QCD systematics of factorizable processes related to DIS.
 
In this Letter, we propose that quantum information measures can be utilized as a novel
probe of the transversity PDFs, and hence of the tensor charges of the nucleon. In particular,
we show that the spin correlations between the scattered electron and the struck quark in
DIS are significantly sensitive to parton-level transverse polarization as quantified
via underlying transversity PDFs. This
sensitivity translates into a direct, calculable mapping between the quantum correlations of
the DIS final state and the tensor charges, $\delta u$ and $\delta d$. Notably, these measures constrain a direction in the
$(\delta u,\,\delta d)$ plane that is {\it distinct} from the isovector
combination, $g_T \equiv \delta u - \delta d$, targeted by lattice QCD and low-energy experiments.
We therefore demonstrate that QI measures supply novel and nontrivial information on particle
interactions in a nonperturbative QCD setting which complement empirical and lattice constraints.

%%%%%%%%%%%%%%%%%%%%%%%%%%%%%%%%%%%%%%%%%%%%%%%%%%%%%%%%%%%%%%%%%%%%%%%%%%%
\noindent\textbf{\textit{Quantum State of} DIS}  --- 
%%%%%%%%%%%%%%%%%%%%%%%%%%%%%%%%%%%%%%%%%%%%%%%%%%%%%%
We consider electron-proton scattering at high momentum transfer, where the DIS subprocess, $e^- + q \to e^- + q$, dominates. The spin degrees of freedom of the final-state electron and struck quark are treated as qubits, so that their joint spin state forms a bipartite quantum system. The quantum state of this system, $\rho_{eq}$, is described by the Fano-Bloch decomposition
\begin{equation}
  \rho_{eq} = \tfrac{1}{4}\left( \mathbb{I}_2\otimes \mathbb{I}_2 + B_i^{e}\,\sigma^i\otimes \mathbb{I}_2 + B_j^{q}\, \mathbb{I}_2\otimes\sigma^j + C_{ij}\,\sigma^i\otimes\sigma^j \right),
    \label{eq:density_matrix}
\end{equation}
where $\mathbb{I}_2$ is the $2\times2$ identity, $\sigma^i$ are the Pauli matrices, $B_i^{e/q}$ describe the spin polarization of the electron/quark, and $C_{ij}$ is the spin-correlation matrix. We emphasize that each of these components is the expectation value of a final-state spin observable, $B_i^{e/q} = \langle \sigma^i \rangle_{e/q}$ and $C_{ij} = \langle \sigma^i \otimes \sigma^j \rangle$, so that $\rho_{eq}$ may be reconstructed directly from the measured single-spin polarizations and spin-spin correlations of the final state. 

The final-state density matrix for the joint electron-quark system is fixed by the DIS scattering amplitude through
\begin{equation}
  (\rho_{eq})_{\alpha\bar\alpha,\alpha'\bar\alpha'} \propto
    \sum_{\lambda\bar\lambda\lambda'\bar\lambda'}
    \mathcal{M}_{\alpha\bar\alpha}^{\lambda\bar\lambda}\ \rho^{e}_{\lambda\lambda'}\,
    \rho^{q}_{\bar\lambda\bar\lambda'}\
    (\mathcal{M}_{\alpha'\bar\alpha'}^{\lambda'\bar\lambda'})^\dagger,
\end{equation}
where
\begin{equation}
  \rho^{e/q} = \frac{I_2 + \vec b^{\,e/q}_\perp\cdot\vec\sigma}{2}
\end{equation}
are the density matrices of the initial electron and quark, and $\mathcal{M}_{\alpha\bar\alpha}^{\lambda\bar\lambda}$ is the hard-scattering amplitude (indexed over helicities), given explicitly in the Supplemental Material. A crucial feature of the $t$-channel photon exchange is that the transition matrix $\mathcal{M}$ is diagonal in the helicity basis: unpolarized or longitudinally polarized beams therefore generate only separable final states, and nontrivial quantum correlations arise when the initial state is prepared as a coherent superposition of helicity eigenstates. Following Ref.~\cite{Cheng:2025zaw}, we take the initial electron and proton to be transversely polarized along a common direction, identified with spin projections along $\sigma_1$; for the final state, we work in the helicity basis, quantized along the outgoing electron direction. Throughout, we fix the transverse polarization of the initial electron beam to $b_\perp^e = 0.7$.

%%%
\noindent\textbf{\textit{Quantum Hierarchy of} DIS} --- 
%%%
The spin state of the final $e^{-}q$ pair is in general a mixed quantum state, featuring a hierarchy of nested and/or complementary quantum correlations. In this section, we outline the quantum information measures of the density matrix considered in our analysis. 

The canonical resource of quantum information theory is entanglement. Given a quantum state, $\rho$, we say that the state is entangled if and only if it cannot be decomposed into a separable state. Concurrence is a measure of entanglement for a two-qubit system: 
\begin{equation}
    \mathcal{C}[\rho]=\max(0,\lambda_1- \lambda_2-\lambda_3-\lambda_4)\ ,
\end{equation}
where \(\{\lambda_i\}\) are the eigenvalues (in decreasing order) of \(\sqrt{\sqrt{\rho}\tilde{\rho}\sqrt{\rho}}\); here, \(\tilde{\rho}=(\sigma_2\otimes\sigma_2)\rho^*(\sigma_2\otimes\sigma_2)\), and \(\rho^*\) is the complex conjugate of \(\rho\). $\mathcal{C}[\rho]=0$ corresponds to a separable state, while $\mathcal{C}[\rho]=1$ is a maximally entangled state, such as a Bell state, {\it e.g.}, $\ket{\Phi}=(\ket{\uparrow\downarrow} + \ket{\downarrow\uparrow})/\sqrt{2}$. For DIS, the concurrence of the $e^{-}q$ final state is maximized in the back-scattering limit, $\cos\theta = -1$, where~\cite{Cheng:2025zaw}
\begin{equation}
    \mathcal{C}[\rho_{eq}] = b_{\perp}^{e}b_{\perp}^{q}\ .
\end{equation}
%%%
Thus, Bell states are produced when $b_{\perp}^{e}=b_{\perp}^{q}=1$.
%%%
\begin{figure}
    \centering
    \includegraphics[width=0.49\linewidth]{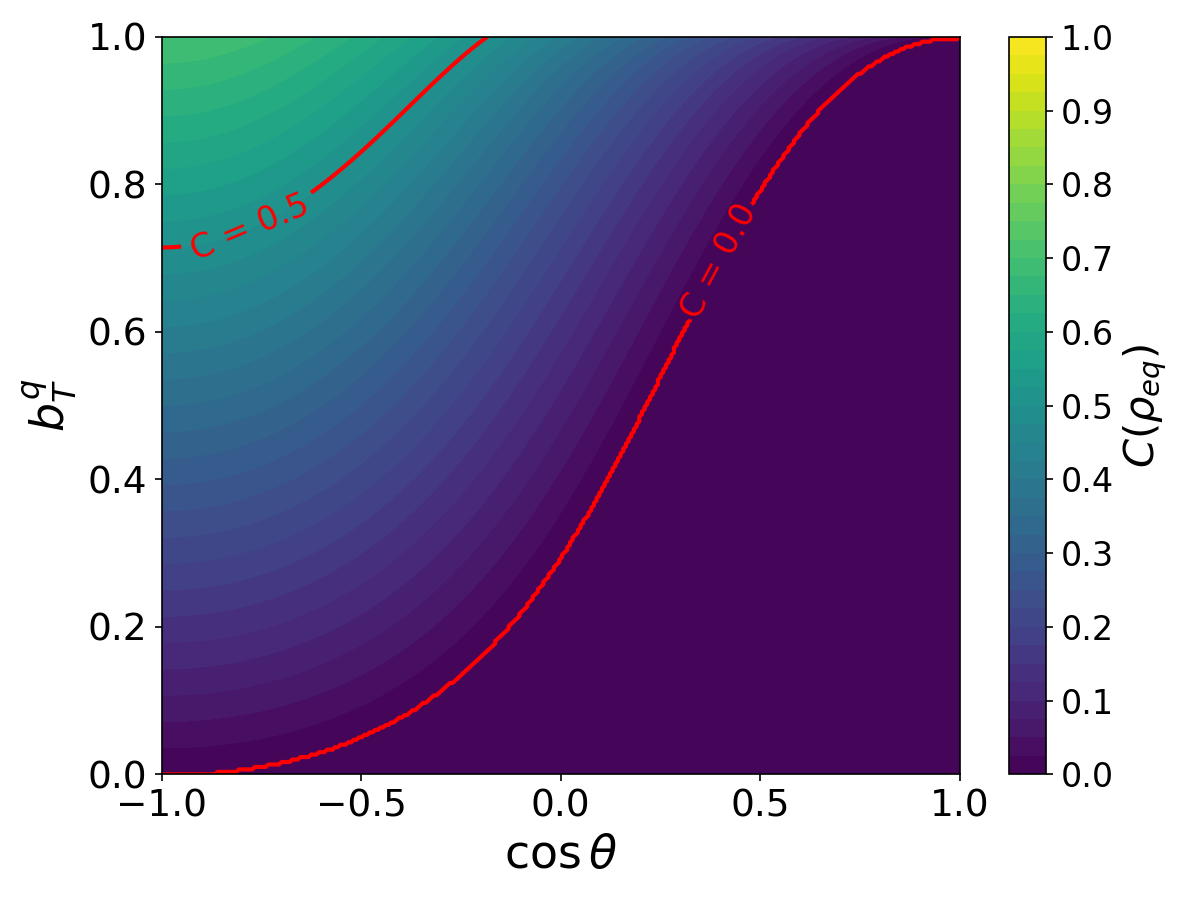}
    \includegraphics[width=0.49\linewidth]{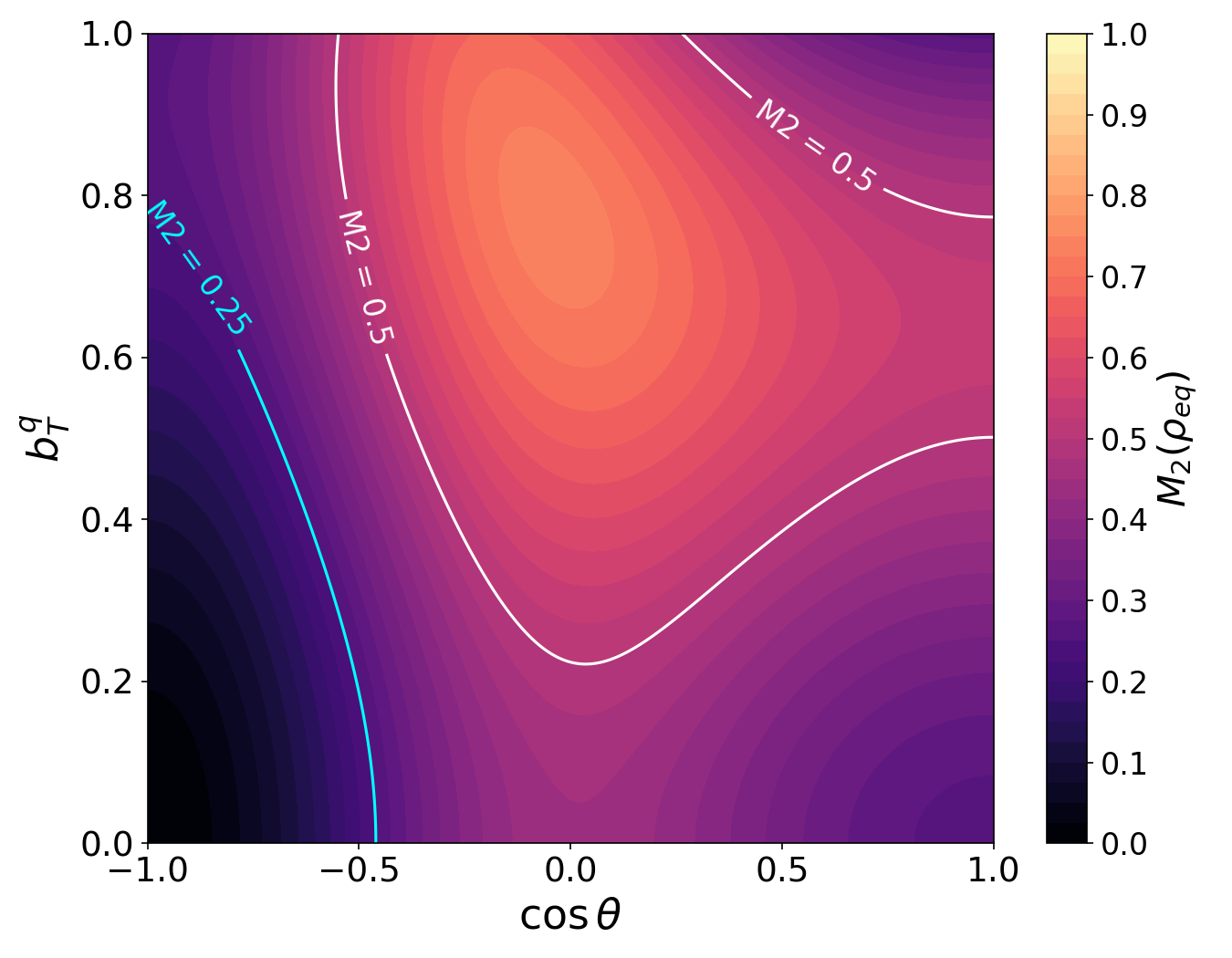}
    \includegraphics[width=1\linewidth]{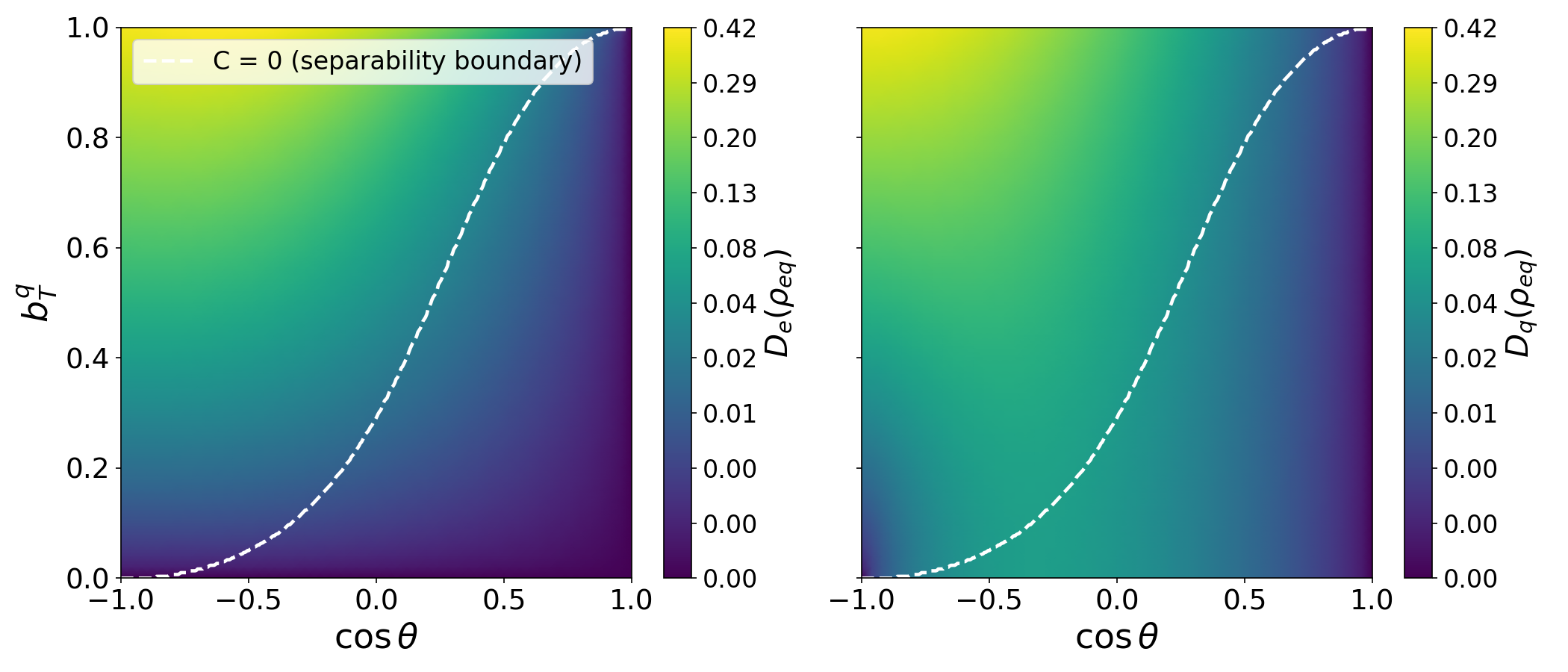}\\
    \includegraphics[width=1\linewidth]{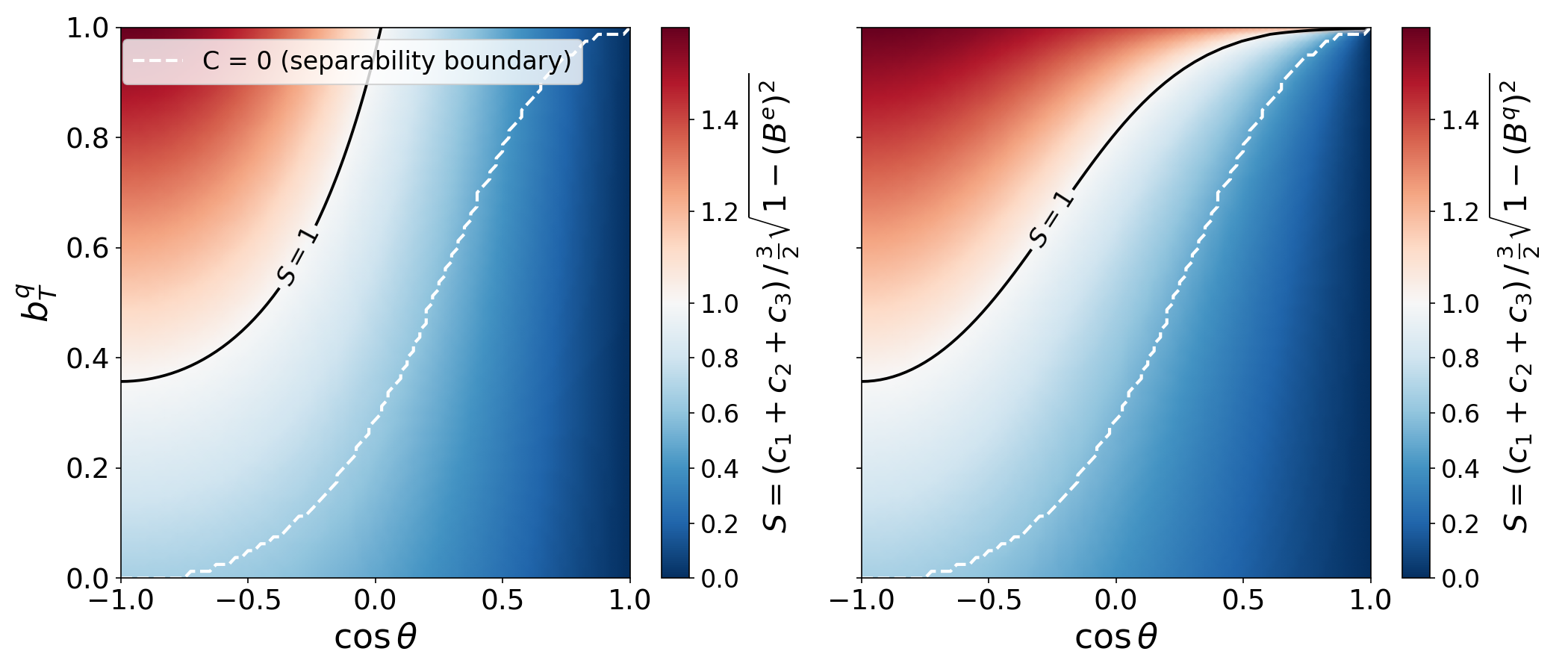}
    \caption{The quantum concurrence (top left); magic (top right); electron and quark discord (middle left and right); and electron and quark steering (bottom left and right). In all panels, we fix $b_{\perp}^{e}=0.7$.}
    \label{fig:quant_hierarchy}
\end{figure}
%%%

In addition to entanglement, quantum magic has become regarded as a crucial quantum resource for quantum computation. Quantum magic measures the failure of a quantum state to be simulated via so-called \textit{stabilizer} techniques~\cite{Gottesman:1998hu}.  We compute magic using the second stabilizer Rényi entropy,
\begin{equation}
    M_2(\rho)=-\log_2\left[\frac{\sum_{P\in\mathcal{P}_n}(\text{Tr}[P\rho])^4}{\sum_{P\in\mathcal{P}_n}(\text{Tr}[P\rho])^2}\right]\ ,
\end{equation}
where 
\begin{equation}
    \mathcal{P}_n=P_1\otimes P_2\otimes \dots \otimes P_n,~P_i\in\{\mathbb{I}_2,\sigma_1,\sigma_2,\sigma_3\}.
\end{equation}
is the set of Pauli strings.

While entanglement and magic are often used as measures of nonclassicality, the true classical-quantum boundary is measured through quantum discord. Discord captures the correlations destroyed by a local measurement of one subsystem, namely the gap between the total correlations of the state and those of a purely classical nature. It vanishes only for classically correlated states, and so registers quantum correlations even where entanglement is absent. In a bipartite system composed of electron, \textit{e}, and quark, \textit{q}, quantum discord for the electron subsystem is defined by
\begin{align}
    D_e(\rho_{eq})&=S(\rho_q)-S(\rho_{eq})\\
                 &\ \ \ \ +\, \min_{\hat{n}}[p_{+\hat{n}}S(\rho_{+\hat{n}})+p_{-\hat{n}}S(\rho_{-\hat{n}})]\ , \nonumber
\end{align}
where

\begin{flalign}
    p_{\pm\hat{n}}=\frac{1\pm \mathbf{\hat{n}}\cdot\mathbf{B}^q}{2}\ ,&\quad
\rho_{\pm\hat{n}}=\frac{\mathbb{I}_2+\mathbf{B}^e_{\pm\mathbf{\hat{n}}}\cdot\vec{\sigma}}{2}\ ,\\
    \mathbf{B}^e_{\pm \mathbf{\hat{n}}}=&\frac{\mathbf{B}^e\pm \mathbf{C}\cdot\mathbf{\hat{n}}}{1\pm\mathbf{\hat{n}}\cdot\mathbf{B}^q}\ .
\end{flalign}
$\rho_{\pm\hat{n}}$ is a conditional quantum state of the electron obtained after a measurement of the quark spin in the $\ket{\pm\hat{n}}$ direction, which has probability $p_{\pm\hat{n}}$. $S(\rho) = -\text{Tr}[\rho\log_{2}(\rho)]$ is the Von Neumann entropy. Discord of the quark subsystem is obtained from similar expressions with \textit{e}$\leftrightarrow$\textit{q}.

Steerability of a two-qubit system indicates the ability to understand the entire ensemble of possible states one qubit can inhabit based solely on knowledge of the choice and outcome of a measurement on the other qubit.  Steerability for the electron subsystem can be measured through the violation of the inequality 
\begin{equation}
    c_1+c_2+c_3\leq\frac{3}{2}\sqrt{1-(\mathbf{B}^e)^2}\ ,
\end{equation}
where the $\{c_i\}$ are the singular values of the matrix
\begin{equation}
    \mathfrak{C}_{jk} \equiv C_{jk}-B^e_jB^q_k\ .
\end{equation}
Quark steerability is obtained through the substitution \textit{e}$\leftrightarrow$\textit{q}. In Fig.~\ref{fig:quant_hierarchy}, we show the quantum hierarchy of correlations produced in DIS at parton level with respect to the scattering angle and the assumed initial quark transverse polarization. In the top panels, we show the symmetric information measures, the concurrence (left) and magic (right). We see that, while entanglement is maximized in the back-scattering region ($\cos \theta \sim -1$), quantum states with significant quantum advantage can occur in the more central scattering region ($\cos \theta \sim 0$).

In the middle panels, we show quantum discord for the final-state electron (left) and quark (right). Both the electron and quark discord show quantum correlations which persist beyond the region of entangled states as identified by the concurrence, delineated by the white-dashed line.
Thus, we find that quantum correlations survive into regions wherein the final state is separable, such that discord remains a sensitive probe even when entanglement is absent. 
Furthermore, we see that the quark discord is asymmetric as compared to the electron. This is the first example where discord-asymmetric states have been studied in a high-energy collider setting.

Finally, in the bottom panels of Fig.~\ref{fig:quant_hierarchy} we show the steering measure of the final-state electron (left) and quark (right). In both panels, steerable states reside in the regions above the solid black line where
\begin{equation}
    \mathcal{S}_{e/q} \equiv \frac{2}{3}\frac{c_{1}+c_{2}+c_{3}}{\sqrt{1-(\mathbf{B}^{e/q})^{2}}}\geq 1\ .
\end{equation}
Although we see that both regions are subsets of the set of entangled states (as they must be), there is again a notable asymmetry between the steerable states of the electron versus those of the quark, with the region of quark-steerable states larger in particular. 

The steerability region holds experimental implications: in principle, the ensemble of possible quark transverse polarizations can be inferred from a measurement on the electron alone. This is especially valuable given that the quark is accessible only through its jet, so that the larger quark-steerable region suggests a potential handle on the quark's quantum state through the measured electron. We plan to explore further this utility of QI for DIS in future work.

%%%
\begin{figure}
    \centering
    \includegraphics[width=1\linewidth]{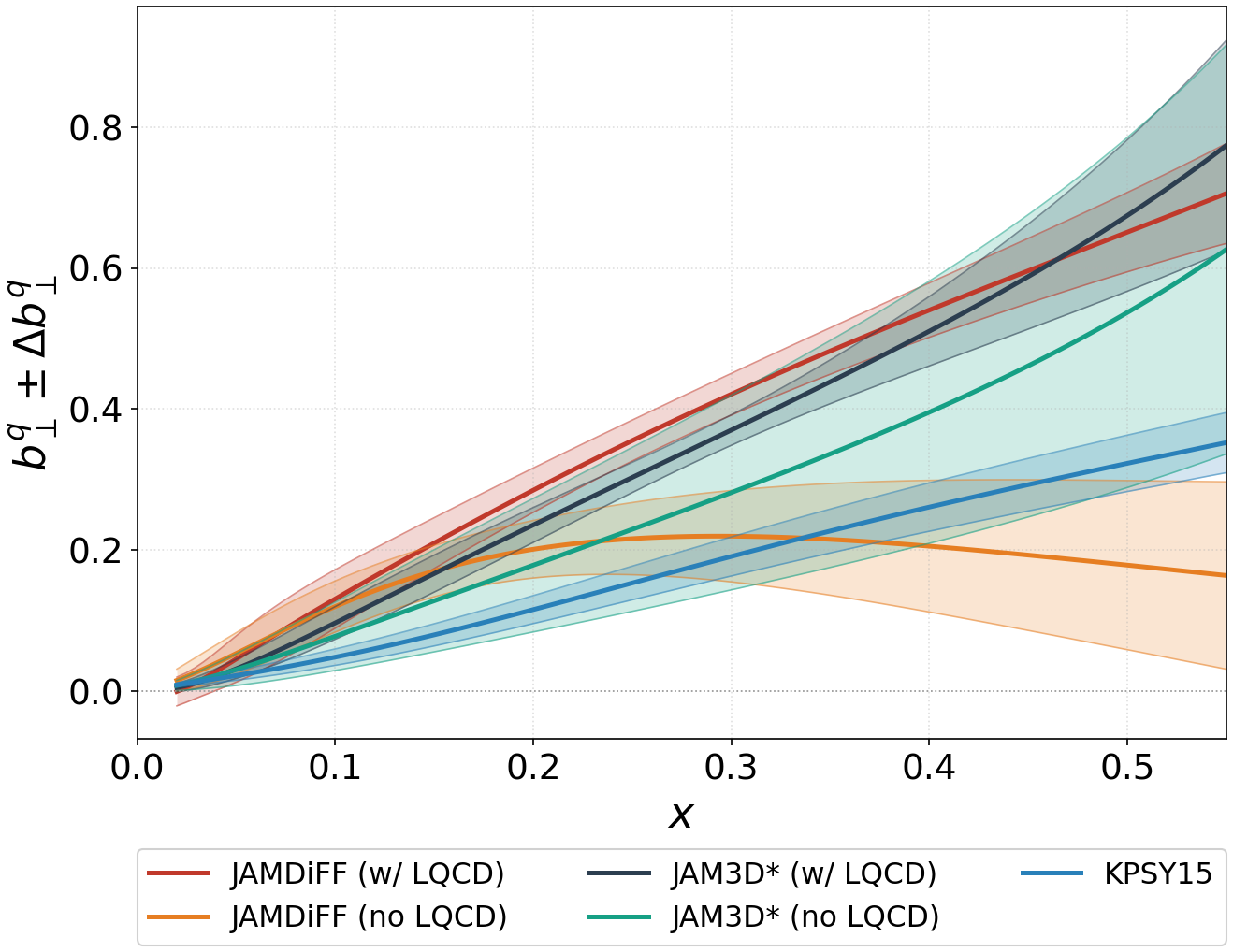}
    \caption{We plot $b_\perp^{\langle q\rangle}$ {\it vs}.~$x$ as given by Eqs.~(\ref{eq:bperp})--(\ref{eq:alpha}) for 5 different transversity PDF analyses or models within their $ 1\sigma$ ranges; we fix $Q=12$ GeV and $b_{\perp}^{p}=0.7$.}
    \label{fig:bqperp}
\end{figure}
%%%
%%%%%%%%%%%%%%%%%%%%%%%%%%%%%%%%%%%%%%%%%%%%%%%%%%%%%%%%%%%%%%%%%%%%%%%%%%
\noindent\textbf{\textit{Connection to Nonperturbative Models}} --- 
%%%%%%%%%%%%%%%%%%%%%%%%%%%%%%%%%%%%%%%%%%%%%%%%%%%%%%%%%%%%%%%%%%%%%%%%%%
%
Knowledge of the DIS quantum state, $\rho_{eq}$, depends upon the underlying
nonperturbative structure of the proton. In particular, the struck (interacting) quark
carries some fraction, $\alpha$, of the transverse polarization of its parent proton,
\begin{equation}
  b_\perp^{\langle q\rangle} = \alpha\, b_\perp^{p},
    \label{eq:bperp}
\end{equation}
where
\begin{equation}
  \alpha(x,Q^2) \;=\; \frac{\sum_q e_q^2\, h_{1,q}(x,Q^2)}{\sum_q e_q^2\, f_q(x,Q^2)}
  \label{eq:alpha}
\end{equation}
defines a light-quark flavor-averaged ratio of the transversity PDFs, $h_{1,q}$, to the unpolarized PDFs,
$f_q$. Here we have included explicit dependence on the quark charges, $e_{q}$, collinear parton fraction, $x$,
and virtuality, $Q^2$. $\alpha$ parametrizes the nonperturbative input entering $\rho_{eq}$.
The quantum correlations of the final state therefore act as a direct proxy for the transverse
polarization carried by the struck quark, and hence for the transversity PDFs themselves.
 
Eq.~(\ref{eq:alpha}) assumes no flavor-tagged measurements at the hadronic level. Each transversity
distribution is weighted by its squared quark charge, $e_q^2$, analogously to the unpolarized
structure function, $F_2 = x\sum_q e_q^2 f_q$, in the quark-parton model. Inclusion of, {\it e.g.}, flavor-tagged final states would
isolate individual quark flavors and shift $\alpha$ --- an aspect we defer to future work. Because the light quarks are indistinguishable in
the inclusive measurement, the spin state is treated as a flavor mixture. The second-generation
quarks contribute to the unpolarized denominator but are constrained to contribute negligibly
to the transversity numerator by the Soffer bound~\cite{DAlesio:2020vtw},
\begin{equation}
  |h_{1,c/s/b}| \le \tfrac{1}{2}\left(f_{c/s/b} + \Delta f_{c/s/b}\right),
\end{equation}
which forces $h_{1,c/s/b}$ to nearly vanish beyond small $x$. Similarly, the partonic scattering angle, $\theta$, is fixed by DIS kinematics involving the parton fraction, $x$, virtuality, $Q^2$, and center-of-mass energy, $s$: $\cos\theta = 1 - 2Q^2/{x s}$.
 
In Fig.~2 we show the $1\sigma$ ranges for the quark transverse polarization,
$b_\perp^{\langle q\rangle}(x)$, based upon five transversity PDF analyses or models. These are: JAMDiFF with (red) and
without (orange) lattice-QCD (LQCD) constraints~\cite{Cocuzza:2023vqs}, JAM3D with (purple) and
without (green) LQCD~\cite{Gamberg:2022kdb}, and KPSY15 (light blue)~\cite{Kang:2015msa}. In computing $\alpha$, we take the unpolarized PDFs from JAM~\cite{Cocuzza:2022hse} for consistency. These
calculations agree well at small $x$ but produce a broader spread at larger values of $x$ as shown.

It is precisely the spread in $b_{\perp}^{\langle q\rangle}(x)$ for $x\! \gtrsim\! 0.4$ to which the hierarchy of quantum measures
is sensitive. As shown in Fig.~3, at a representative kinematic point ($x=0.5$, $Q=12$~GeV,
$\sqrt{s}=20$~GeV) the five analyses yield clearly distinct values of concurrence (upper left), quark discord (upper right),
quark steering (lower left), and magic (lower right), so that the quantum hierarchy discriminates between competing transversity
models. Crucially, we find that the requirement of quark steerability, marked by the red dashed line in the lower left panel, defines a strong discriminator between transversity scenarios with or without LQCD constraints, with the former consistently quark-steerable at the 1$\sigma$ level, unlike the latter.

\begin{figure}
    \centering
    \includegraphics[width=0.49\linewidth]{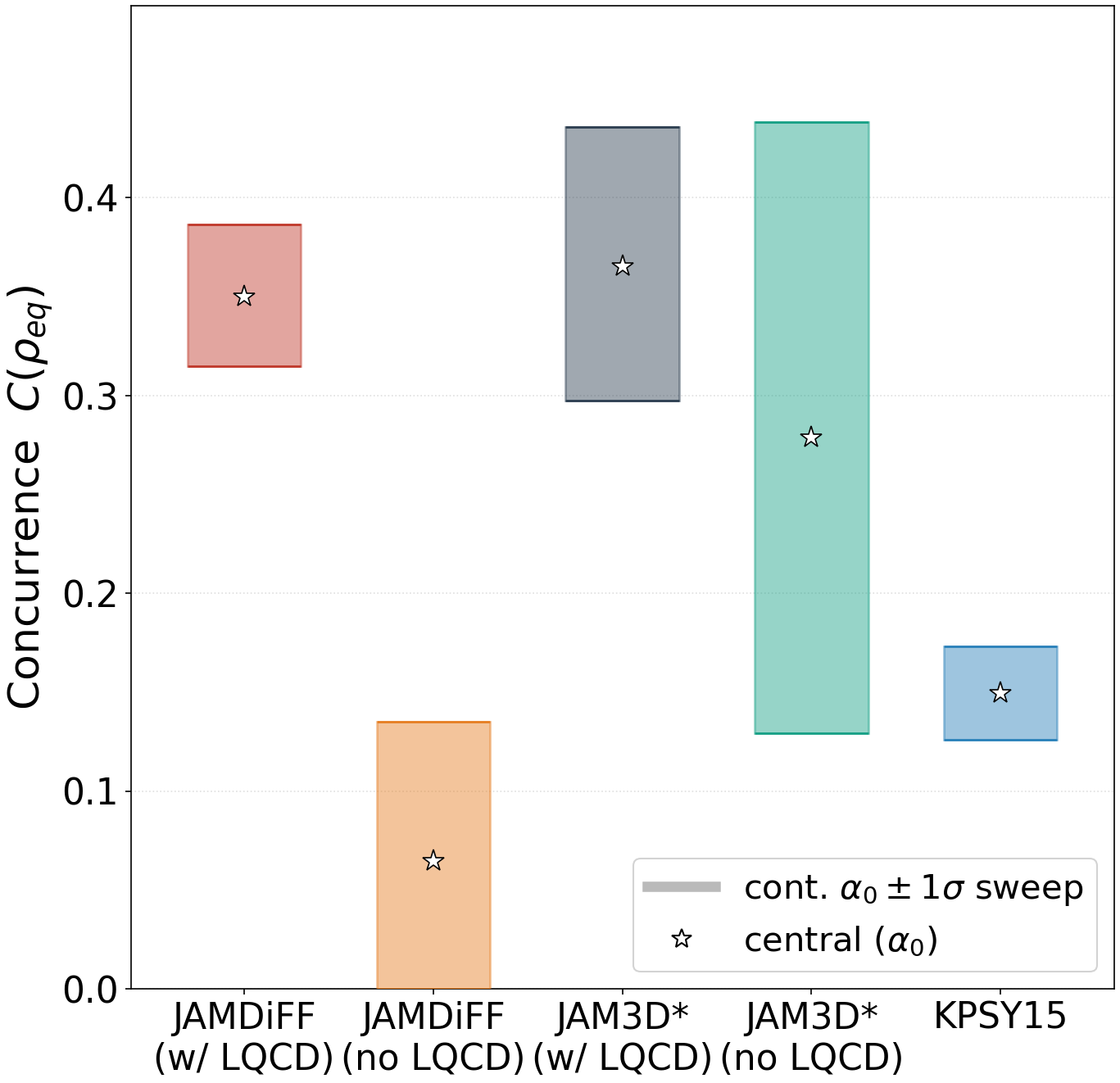}
    \includegraphics[width=0.49\linewidth]{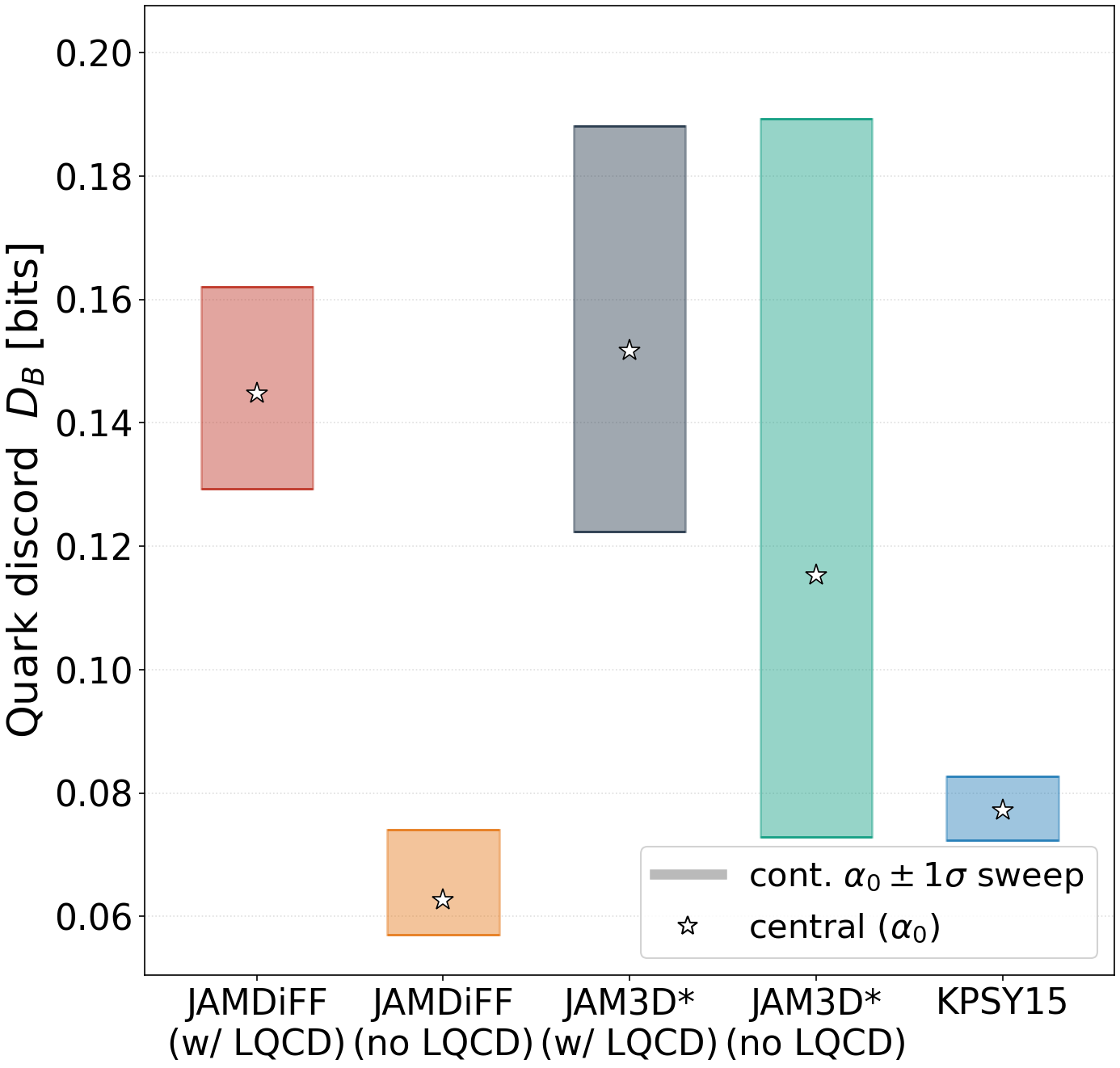}\\
    \includegraphics[width=0.49\linewidth]{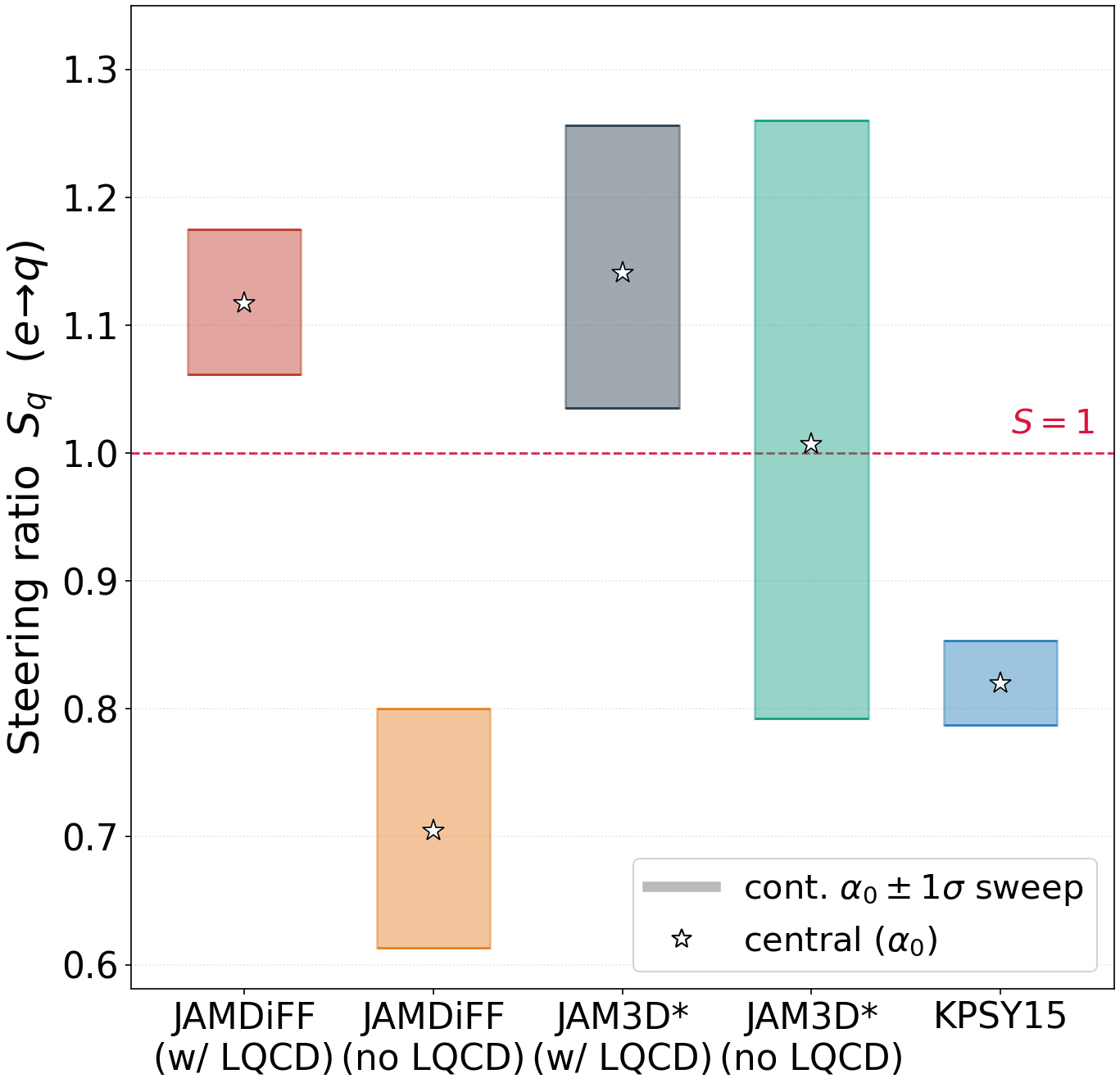}
    \includegraphics[width=0.49\linewidth]{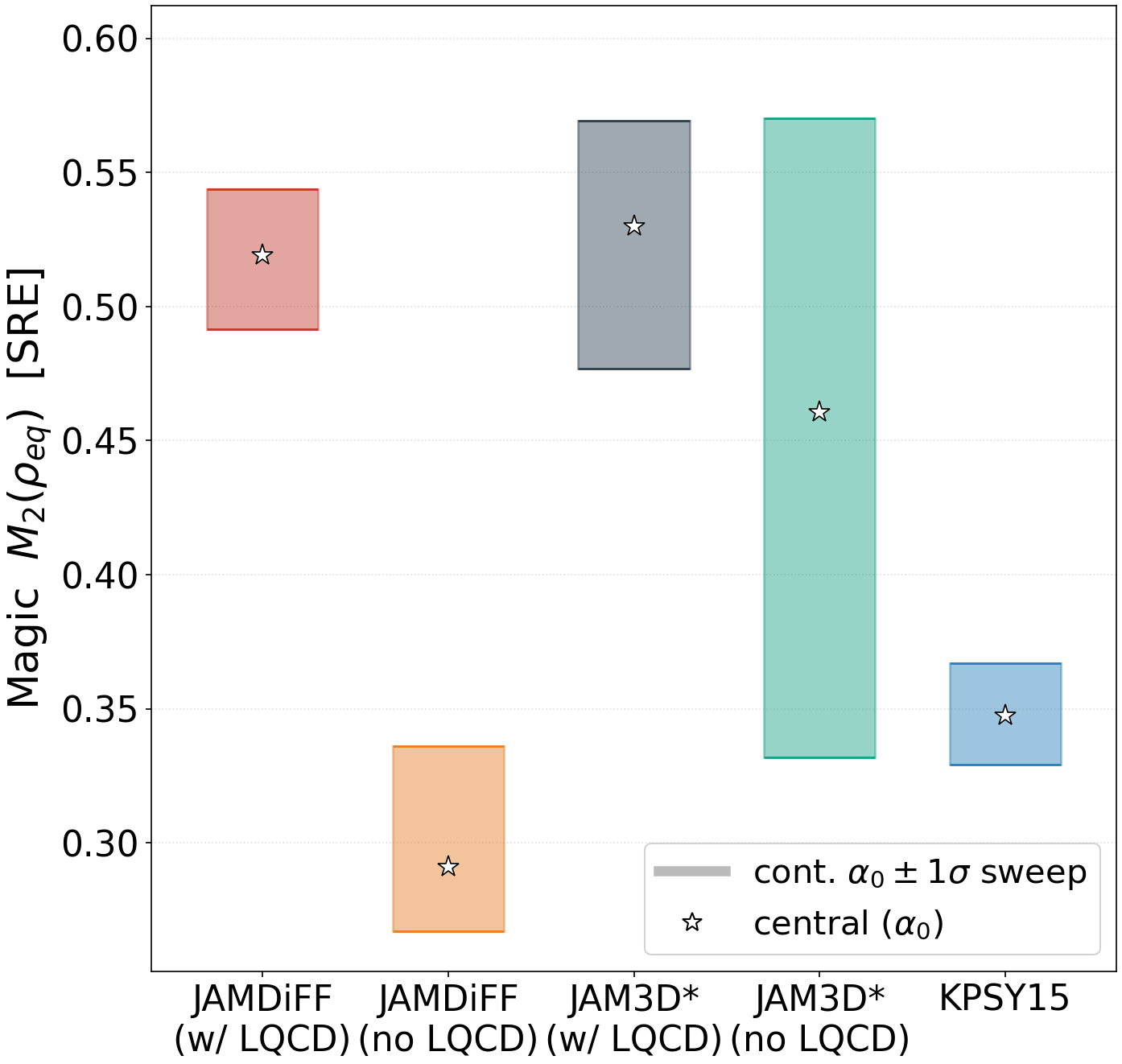}
    \caption{The quantum concurrence, discord, steering, and magic at fixed $x=0.5$, $Q=12$ GeV, and $\sqrt{s}=20$ GeV, shown for 5 different transversity analyses; $b_{\perp}^{e}=b_{\perp}^{p}=0.7$ are fixed.}
    \label{fig:ModelDepConcurrence}
\end{figure}

%%%%%%%%%%%%%%%%%%%%%%%%%%%%%%%%%%%%%%%%%%%%%%%%%%%%%%%%%%%%%%%%%%%%%%%%%%
\noindent\textbf{\textit{Constraints on Tensor Charges and BSM Implications}} --- The sensitivity of the quantum hierarchy to the underlying transversity presents a compelling
question regarding the utility of quantum information to inform hadronic substructure. Yet the
transversity PDFs are themselves quasi-derived, model-dependent objects whose relative
inaccessibility obfuscates a clean interpretation. To crystalize the potential of QI, we
therefore connect these quantum variables not to the transversity distributions directly, but
to the tensor charges of the up and down quarks,
\begin{equation}
  \delta q\, (Q^2) = \int_0^1 dx\, \left[\, h_{1,q}(x,Q^2) - h_{1,\bar q}(x,Q^2) \,\right],
\label{eq:tensor_charges}
\end{equation}
which are fundamental charges of the nucleon. 
 
\begin{figure*}[t]
     \centering
    \includegraphics[width=0.45\linewidth]{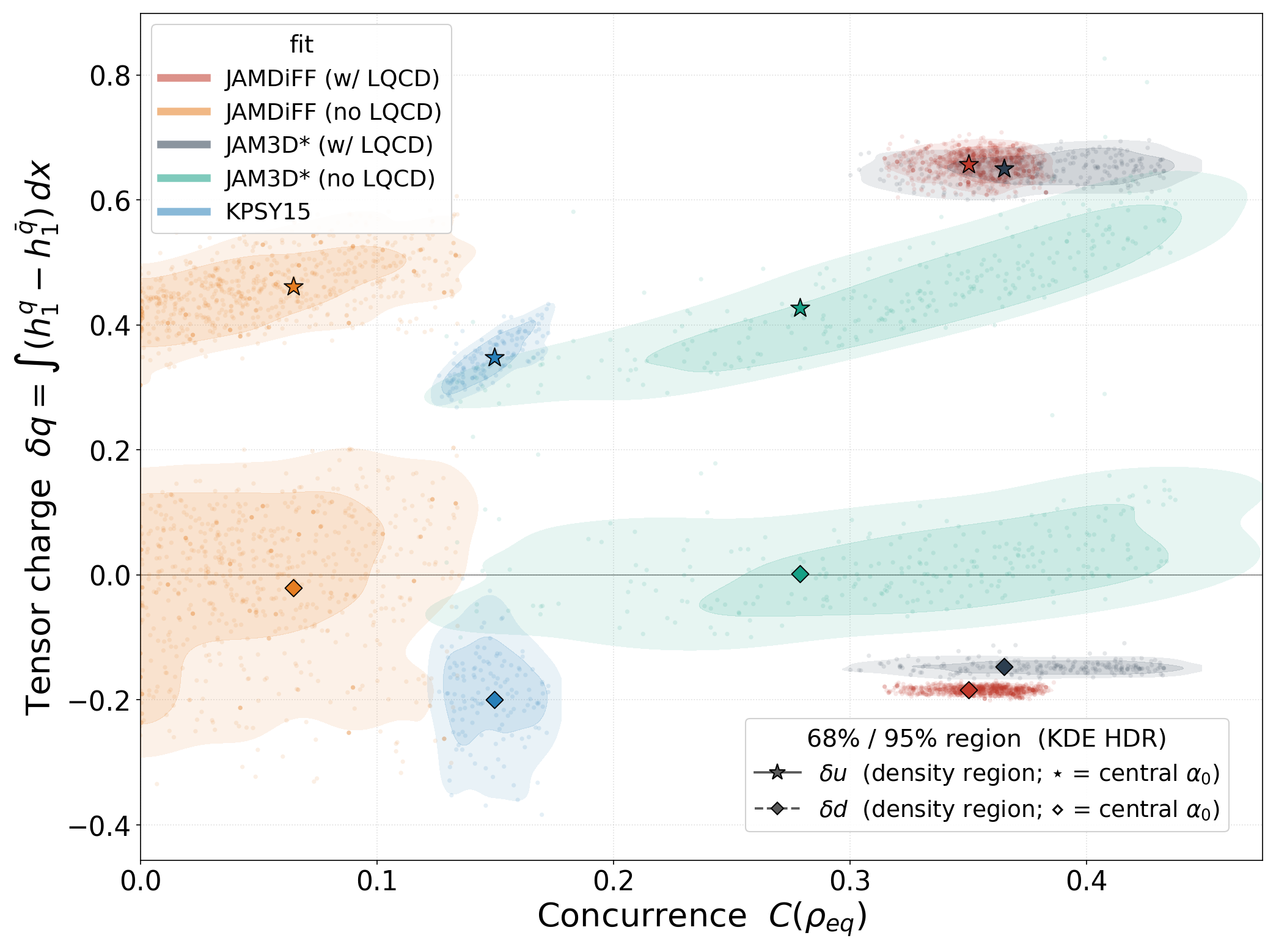}
    \includegraphics[width=0.45\linewidth]{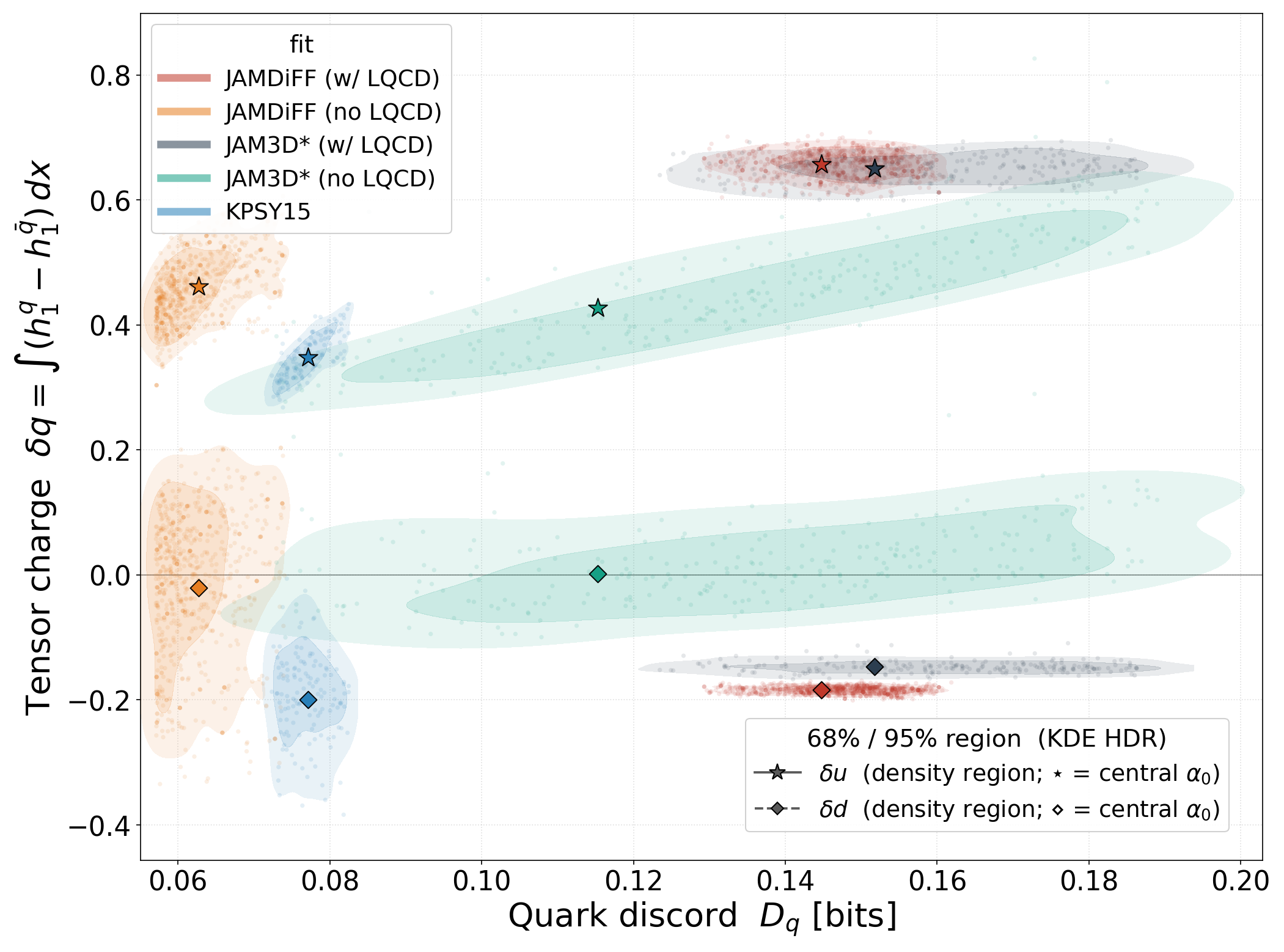}\\
    \includegraphics[width=0.45\linewidth]{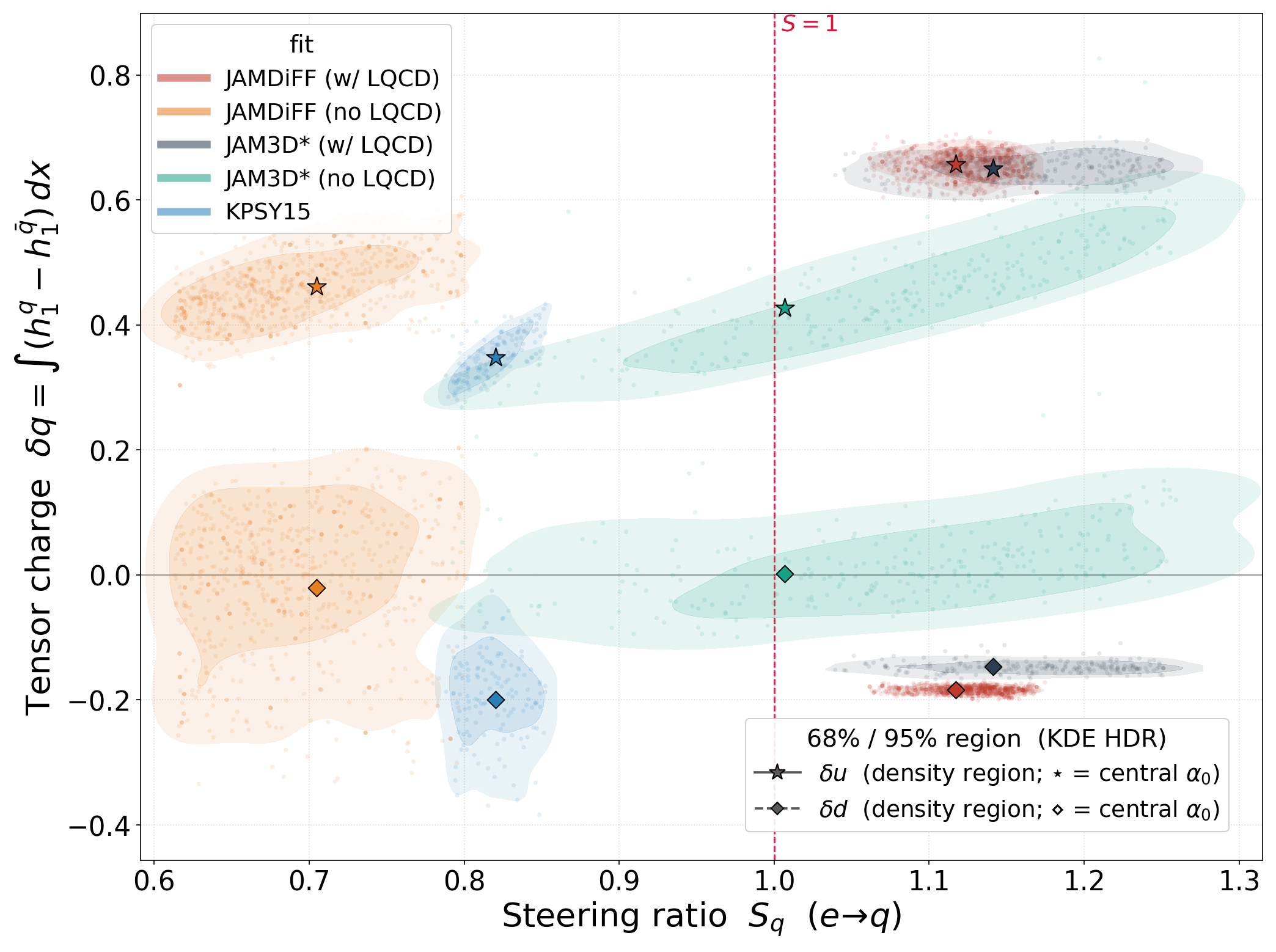}
    \includegraphics[width=0.45\linewidth]{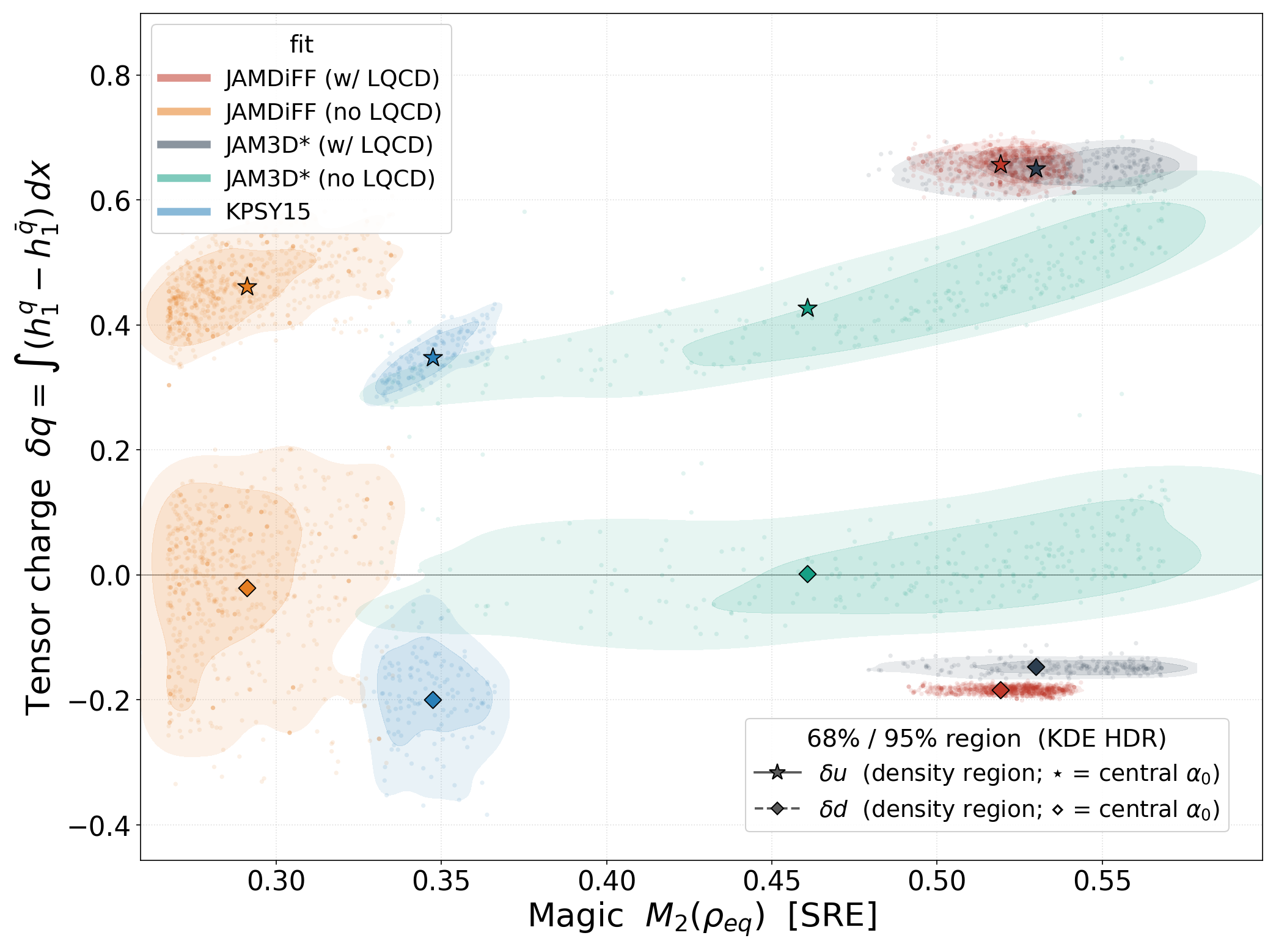}
    \caption{The up- and down-quark tensor charges ($\delta u,\, \delta d$) for 5 different transversity analyses at $x=0.5$, $Q=12$ GeV, plotted against the corresponding values of the quantum concurrence, quark discord, magic, and steering of the quark.}
    \label{fig:QITensor}
 \end{figure*}

Our main proposal stems from the fact that the tensor charges and the quantum hierarchy are simultaneously
functionals of the same object. A given transversity fit predicts a definite range of tensor
charges and a definite range of quantum correlations. To make this connection
concrete, we propagate the transversity replicas within the $1\sigma$ bands of Fig.~\ref{fig:bqperp}
through both the moment integral of Eq.~(\ref{eq:tensor_charges}) and the quantum measures evaluated at the
reference kinematics of Fig.~\ref{fig:ModelDepConcurrence}. Since both are inherited from the same set of $h_{1,q}$, each fit populates
a correlated region in the plane of a quantum measure against a tensor charge, which we
display in Fig.~\ref{fig:QITensor}.

The reconstruction of the
spin-correlation measures, {\it i.e.}, quantum tomography of $\rho_{eq}$, is tied directly to the predicted tensor charges. Each fit of the transversity PDFs maps onto a definite locus in the plane of a quantum measure against $\delta u$ or $\delta d$.
Read in the direction of measurement, a tomographic determination of any of these measures
therefore selects a band of allowed tensor charges. In this sense, performing quantum tomography
in DIS directly informs the allowed values of $\delta u$ and $\delta d$. We stress
that this inference proceeds entirely from invariants of the measured density matrix, and is
thus distinct from and complementary to the nonperturbative model assumptions and systematics that underlie conventional global
analyses.
 
We note in particular that the threshold nature of steering renders this connection especially decisive. Because a quantum state
either is or is not steerable, establishing a quark-steerable final state at fixed kinematics selects the region of allowed tensor-charges consistent with the measurement. Notably, we find that this binary criterion is already
sufficient to distinguish the quark tensor charges extracted with and without lattice-QCD input. We see that the transversity fits informed by lattice data imply fully quark-steerable states at $1\sigma$, whereas those without
lattice constraints do not. The observation of quark steerability in DIS can thereby furnish
a sharp, tomographically clean measure which can act as a discriminant between competing determinations of the proton's
tensor charges.

Lastly, we note that our findings carry BSM implications for the interpretation of neutron EDM~\cite{Chupp:2017rkp} or
nonstandard interaction searches in precision $\beta$-decay. For the former, the EDM of the neutron goes as
\begin{equation}
    d_n = \sum_q\ \delta q^n\, d_q\ ,
\end{equation}
with the quark-level EDMs, $d_q$, induced by dim-6 dipole operators in Standard Model Effective Field Theory (SMEFT)~\cite{Grzadkowski:2010es} and
weighted by the flavor-dependent quark tensor charges in the neutron, $\delta q^n$, which are related to those of the proton by isospin; crucially, while $\beta$-decay is generally sensitive to the
isovector flavor combination, $g_T = \delta u - \delta d$, the quantum hierarchy constraints shown above, especially that associated with the steerability, impact a distinct flavor combination given the definition of Eq.~(\ref{eq:alpha}).
We therefore conclude that information on quantum correlations has the potential to relieve degeneracies in SMEFT-based BSM
analyses, in addition to its sensitivity to nonperturbative QCD.

%%%%%%%%%%%%%%%%%%%%%%%%%%%%%%%%%%%%%%%%%%%%%%%%%%%%%%%%%%%%%%%%%%%%%%%%%%
\noindent\textbf{\textit{Conclusions}} --- In this Letter, we have proposed the hierarchy of quantum correlations produced in
DIS as a novel probe of the proton's nonperturbative structure. Treating the scattered electron and struck quark as a bipartite qubit system, we showed that the concurrence, quantum discord, steering, and magic of the final state are each governed by the transverse polarization carried by the struck quark within the proton, and hence by the transversity PDFs. Because these measures are invariants of the spin-density matrix, their reconstruction constitutes a quantum tomography of the DIS final state that constrains the transversity, and thereby the tensor charges, $\delta u$ and $\delta d$. We further found this connection to be sharp enough to provide discriminating power over modern transversity fits in a manner which complements empirical and lattice constraints. Crucially, the establishment of a quark-steerable final state particularly supplies a binary and tomographically clean criterion with the potential to separate determinations of the tensor charges made with and without lattice-QCD input.
 
This program underscores the value of transversely polarized beams together with spin-sensitive measurements of DIS final states at facilities like the EIC, with the quark spin accessed through its fragmentation into a jet and the electron spin through dedicated polarimetry. Natural extensions include a data-driven analysis incorporating realistic fragmentation and detector effects, the treatment of scale evolution and higher-order corrections, and the use of flavor-tagged final states to access the individual quark tensor charges beyond the charge-weighted combination probed here. More broadly, that a fundamental charge of the nucleon can be constrained through the quantum
correlations of a scattering final state suggests that quantum information may serve as a genuinely complementary tool in the precision study of hadronic structure, with implications for both lattice-QCD benchmarks and searches for BSM physics tied to the tensor charge.

\section*{Acknowledgments}
The work of HB and TJH at Argonne National Laboratory was supported by the U.S.~Department of Energy under
contract DE-AC02-06CH11357.
The work of NM is supported in part by the U.S.~Department of Energy under grant No.~DEFG02-13ER41976/DE-SC0009913, and the DOE QuantISED program through the theory consortium ``Intersections of QIS and Theoretical Particle Physics'' at Fermilab (FNAL 20-17) under contract No.~89243024CSC000002. NM also acknowledges the hospitality of Argonne National Laboratory and the University of Chicago where part of this research was performed.

\bibliography{ref}
\newpage
% --- SUPPLEMENTARY MATERIAL ---
\onecolumngrid
\begin{center}
\textbf{\large Supplemental Material}
\end{center}

% Reset equation, figure, table, and page counters
\setcounter{equation}{0}
\setcounter{figure}{0}
\setcounter{table}{0}
\setcounter{page}{1}
\makeatletter
\renewcommand{\theequation}{S\arabic{equation}}
\renewcommand{\thefigure}{S\arabic{figure}}
\renewcommand{\thetable}{S\arabic{table}}
\renewcommand{\bibnumfmt}[1]{[S#1]}

\section{Density Matrix for Deep Inelastic Scattering}
%Hard partonic scattering for DIS at tree level features only a single diagrammatic contribution to the cross section: $t$-channel exchange between the electron and struck quark via a virtual photon.  At this energy scale, QED-like interactions dominate the scattering process, and thus contributions from Z-exchange are excluded.  In the massless limit for both electron and quark, helicity conservation ensures only four nonzero matrix elements exist in the helicity basis:
%
In the Letter above, we organize our calculations about DIS kinematics ($x=0.5$, $Q=12$ GeV) wherein amplitudes are dominated by pure photon exchange and in which the leading-twist, massless limit is applicable with only weak corrections. With these assumptions, the relevant parton-level amplitudes at leading order in the helicity basis are given by:
\begin{subequations}
    \begin{align}
        \mathcal{M}_{\uparrow\Uparrow}^{\uparrow\Uparrow}&\approx2e_qe^2\frac{\hat{s}+t}{t}e^{i\phi}\ ,\\
        \mathcal{M}_{\uparrow\Downarrow}^{\uparrow\Downarrow}=\mathcal{M}_{\downarrow\Uparrow}^{\downarrow\Uparrow}&\approx 2e_qe^2\frac{\hat{s}}{t}\ ,\\
        \mathcal{M}_{\downarrow\Downarrow}^{\downarrow\Downarrow}&\approx 2e_qe^2\frac{\hat{s}+t}{t}e^{-i\phi}\ ,
    \end{align}
\end{subequations}
where upper and lower indices $\uparrow/\downarrow~,~\Uparrow/\Downarrow$ indicate incoming and outgoing spin states of the electron and quark, respectively.  $e_q$ indicates the quark charge, to be removed later from the normalization of density matrix, and $\hat{s}=xs$ is the partonic invariant mass of the DIS interaction, while $t=-Q^2$ is the DIS virtuality common to both the parton-level and hadronic processes. We also note the dependence below on the azimuthal scattering angle, $\phi$, which we set to $\phi\!=\!0$ in the calculations presented in this Letter, but leave general in the formalism below. From the amplitudes above, we assemble the transition matrix,
\begin{equation}\label{eq:TransitionMatrix}
    \mathcal{M}=
         2e_qe^2
        \begin{pmatrix}
            \frac{\hat{s}+t}{t}e^{i\phi} & & & \\
            & \frac{\hat{s}}{t} & & \\
            & & \frac{\hat{s}}{t} & \\
            & & & \frac{\hat{s}+t}{t}e^{-i\phi}
        \end{pmatrix}\ .
\end{equation}
In this system, the expression for the density matrix greatly simplifies to 
\begin{equation}
    \rho_{eq}\propto \mathcal{M}(\rho^e\otimes\rho^q)\mathcal{M}^\dagger\ .
\end{equation}
The fully normalized joint density matrix is thus
\begin{equation}
    \rho_{eq}=
    \frac{1}{2(1+r^2)}\begin{pmatrix}
        r^2 & rb_\perp^qe^{i\phi} & rb_\perp^ee^{i\phi} & r^2 b_\perp^e b_\perp^q e^{2i\phi}\\
        rb_\perp^qe^{-i\phi} & 1 & b_\perp^eb_\perp^q & rb_\perp^ee^{i\phi}\\
        rb_\perp^ee^{-i\phi} & b_\perp^e b_\perp^q & 1 & rb_\perp^q e^{i\phi}\\
        r^2 b_\perp^e b_\perp^q e^{-2i\phi} & rb_\perp^ee^{-i\phi} & rb_\perp^qe^{-i\phi} & r^2
    \end{pmatrix}\ ,
\end{equation}
where \(r\equiv \frac{\hat{s}+t}{\hat{s}}=\frac{\hat{s}-Q^2}{\hat{s}}=\frac{1+\cos\theta}{2}\).  Expressed as a full function of incoming polarizations $(b_\perp^e,b_\perp^q)$ and outgoing scattering angles $(\theta,\phi)$, this density matrix is 
\begin{equation}
    \rho_{eq}[b_\perp^e,b_\perp^q,\theta,\phi]=\frac{1}{4}\begin{pmatrix}
        1-\frac{(1-c_\theta)(3+c_\theta)}{4+(1+c_\theta)^2} & \frac{4b_\perp^q(1+c_\theta)}{4+(1+c_\theta)^2}e^{i\phi} & \frac{4b_\perp^e(1+c_\theta)}{4+(1+c_\theta)^2}e^{i\phi} & 
        \frac{2b_\perp^e b_\perp^q (1+c_\theta)^2}{4+(1+c_\theta)^2}e^{2i\phi} \\
        \frac{4b_\perp^q(1+c_\theta)}{4+(1+c_\theta)^2}e^{-i\phi} & 
        1+\frac{(1-c_\theta)(3+c_\theta)}{4+(1+c_\theta)^2} &
        \frac{8b_\perp^e b_\perp^q}{4+(1+c_\theta)^2} & 
        \frac{4b_\perp^e(1+c_\theta)}{4+(1+c_\theta)^2}e^{i\phi} \\
        \frac{4b_\perp^e(1+c_\theta)}{4+(1+c_\theta)^2}e^{-i\phi} & 
        \frac{8b_\perp^e b_\perp^q}{4+(1+c_\theta)^2} & 
        1+\frac{(1-c_\theta)(3+c_\theta)}{4+(1+c_\theta)^2} &
        \frac{4b_\perp^q(1+c_\theta)}{4+(1+c_\theta)^2}e^{i\phi} \\
        \frac{2b_\perp^e b_\perp^q (1+c_\theta)^2}{4+(1+c_\theta)^2}e^{-2i\phi} & 
        \frac{4b_\perp^e(1+c_\theta)}{4+(1+c_\theta)^2}e^{-i\phi} & 
        \frac{4b_\perp^q(1+c_\theta)}{4+(1+c_\theta)^2}e^{-i\phi} &
        1-\frac{(1-c_\theta)(3+c_\theta)}{4+(1+c_\theta)^2}
    \end{pmatrix}\ .
\end{equation}

%which is a convenient form to study partonic level quantum information theory agnostic to PDF dynamics and kinematic exchange. 

From the decomposition of a bipartite density matrix, we can extract the Bloch polarization vectors and spin correlation matrix by taking the traces of the density matrix projected onto the corresponding basis element of the decomposition:
\begin{subequations}
    \begin{align}
        \mathbf{B}_i^e=\text{Tr}[\rho_{eq}(\sigma_i\otimes \mathbb{I})]&=\begin{pmatrix}
        \frac{4b_\perp^e(1+c_\theta)}{4+(1+c_\theta)^2}c_\phi, & - \frac{4b_\perp^e(1+c_\theta)}{4+(1+c_\theta)^2}s_\phi, & 0
    \end{pmatrix}\\  
        \mathbf{B}_j^q=\text{Tr}[\rho_{eq}(\mathbb{I}\otimes\sigma_j)]&=\begin{pmatrix}
        \frac{4b_\perp^q(1+c_\theta)}{4+(1+c_\theta)^2}c_\phi, & - \frac{4b_\perp^q(1+c_\theta)}{4+(1+c_\theta)^2}s_\phi, & 0
    \end{pmatrix}\\
        C_{ij}=\text{Tr}[\rho_{eq}(\sigma_i\otimes\sigma_j)]&=\begin{pmatrix}
        b_\perp^eb_\perp^q\frac{(4+(1+c_\theta)^2c_{2\phi})}{4+(1+c_\theta)^2} & -b_\perp^eb_\perp^q\frac{(1+c_\theta)^2s_{2\phi}}{4+(1+c_\theta)^2} & 0 \\
        -b_\perp^eb_\perp^q\frac{(1+c_\theta)^2s_{2\phi}}{4+(1+c_\theta)^2} & b_\perp^eb_\perp^q\frac{(4-(1+c_\theta)^2c_{2\phi})}{4+(1+c_\theta)^2} & 0\\
        0 & 0 & -\frac{(1-c_\theta)(3+c_\theta)}{4+(1+c_\theta)^2}
    \end{pmatrix}\ .
    \end{align}
\end{subequations}

\end{document}